\documentclass{iopart}
\usepackage{amsmath,bbm,graphicx,hyperref}
\newcommand{\smx}{{\scriptstyle X}}
\begin{document}
\title{About the probability distribution of a quantity with given
  mean and variance} \author{Stefano Olivares
  \footnote{stefano.olivares@ts.infn.it}} \address{ Dipartimento di
  Fisica, Universit\`a degli Studi di Trieste, I-34151 Trieste, Italy\\
  Dipartimento di Fisica, Universit\`a degli Studi di Milano,
  I-20133 Milano, Italy\\
  CNISM, UdR Milano Statale, I-20133 Milano, Italy}
\author{Matteo G.~A.~Paris \footnote{matteo.paris@fisica.unimi.it}}
\address{Dipartimento di Fisica, Universit\`a degli Studi di Milano,
  I-20133 Milano, Italy \\ CNISM, UdR Milano Statale, I-20133 Milano,
  Italy\\
  INRIM, Strada delle Cacce 91, 10135 Torino, Italy.
  } \date{\today}
%%%%%%%%%%%%%%%%%%%%%%%%%%%
\begin{abstract}
  Supplement 1 to GUM (GUM-S1) recommends the use of maximum entropy   principle (MaxEnt) in determining the probability distribution of a   quantity having specified properties, e.g., specified central   moments. When we only know the mean value and the variance of a   variable, GUM-S1 prescribes a Gaussian probability distribution for   that variable.  When further information is available, in the form   of a finite interval in which the variable is known to lie, we   indicate how the distribution for the variable in this case can be   obtained.  A Gaussian distribution should only be used in this case   when the standard deviation is small compared to the range of   variation (the length of the interval). In general, when the   interval is finite, the parameters of the distribution should be   evaluated numerically, as suggested by I.~Lira [{\em Metrologia},   2009, {\bf 46}, L27]. Here we note that the knowledge of the range   of variation is equivalent to a bias of the distribution toward a   flat distribution in that range, and the principle of minimum   Kullback entropy (mKE) should be used in the derivation of the   probability distribution rather than the MaxEnt, thus leading to an   exponential distribution with non Gaussian features. Furthermore, up   to evaluating the distribution negentropy, we quantify the deviation   of mKE distributions from MaxEnt ones and, thus, we rigorously   justify the use of GUM-S1 recommendation also if we have further   information on the range of variation of a quantity, namely,   provided that its standard uncertainty is sufficiently small   compared to the range.
\end{abstract} \maketitle
%%%%%%%%%%%%%%%%%%%%%%%%%%%%
Supplement 1 to GUM (GUM-S1) \cite{sGUM1} provides assignments of probability density functions for some common circumstances. In particular, it is stated that if we know only the mean value $\bar{x}$ and the variance $\sigma^2_\smx$ of a certain quantity $X$, we should assign a Gaussian probability distribution to that quantity, according to the principle of maximum entropy (MaxEnt) \cite{maxent:57,W87}. The derivation is quite simple, as one has to look for the distribution $p(x)$ maximizing the Shannon entropy: \begin{align}\label{shan} S[p]   = -\int_{\mathbbm R} \!\! dx\, p(x) \log p(x)\,, \end{align} which is given by:
\begin{align}\label{expf} p(x) = \exp\{-\lambda_0 - \lambda_1 x -
  \lambda_2 x^2\}\,, \end{align} where the values of the coefficients
$\lambda_k$ should be determined to satisfy the constraints:
\begin{align} \int_{\mathbbm R} \!\! dx\, p(x)\,x^k = M_k\,,
  \label{constraints} \end{align} with: \begin{align} M_0=1,\quad M_1
  = \bar{x},\quad M_2 = \sigma_\smx^2 + \bar{x}^2
  \label{constraints:M}\,. \end{align} \par However, sometimes we also
know the range of the possible values of the quantity $X$. Two
relevant examples are given by the phase-shift in interferometry,
which is topologically confined in a $2\pi$-window, and by the
displacement amplitude of a harmonic oscillator, whose range of
variation is dictated by energy constraints. In this case, it has been
noticed by I.~Lira in \cite{Lira09} that a Gaussian probability
distribution with support on the real axis can be rigorously justified
only if the standard uncertainty is sufficiently small with respect to
the range of variation of the quantity. More in details, if we have
any information about the range of variation, then this information
should be employed in deriving the distribution maximizing the entropy
as well as in evaluating the values of the coefficients $\{\lambda_0,
\lambda_1, \lambda_2\}$ of the distribution. \par Let us denote
${\mathbbm B} \subset {\mathbbm R}$ the range of the quantity $X$,
i.e., the subset of the real line where the values of $X$ have nonzero
probability to occur. The functional form of the distribution is still
given by the exponential function in Eq.~(\ref{expf}), however with
nonzero support only in ${\mathbbm B}$, whereas the coefficients are
to be determined by formulas like those in Eq.~(\ref{constraints}),
again with $\mathbbm R$ replaced by $\mathbbm B$. It then follows,
e.g., that for a variable which is known {\em a priori} to lie in a
given interval, the maximum entropy distribution is not Gaussian, and
the Gaussian approximation may be employed only if the standard
deviation is small compared to range of the possible values of the
quantity. \par Here we point out that having information about the
range of variation may be expressed as a bias of the distribution
toward a flat distribution in that range and the reasoning presented
in \cite{Lira09} may be subsumed by the minimum Kullback entropy
principle (mKE) \cite{k1,k2,mke1}. The Kullback entropy, or relative
entropy, or Kullback-Leibler divergence, of two distributions $p(x)$
and $q(x)$ reads: \begin{equation} K[p|q] = \int_{\mathbbm R} \!\!
  dx\, p(x) \log\left[ p(x)/q(x)\right]. \end{equation} According to
the mKE, in order to find the distribution $p(x)$ given a bias toward
$q(x)$, we should minimize the function: \begin{equation} {\cal K}[p]
  = K[p|q] + \sum_{k=0}^{2} \lambda_{k}\left[ \int_{\mathbbm R} \!\!
    dx\, p(x)\,x^k - M_k \right], \end{equation} with respect to the
function $p(x)$, obtaining: \begin{equation} p(x) = q(x)
  \exp\{-\lambda_0 - \lambda_1 x - \lambda_2 x^2\}, \label{mKE:sol}
\end{equation} where the parameters $\lambda_k$ can be still
(numerically) computed by using Eq.~(\ref{constraints}).
Eq.~(\ref{mKE:sol}) represents the probability distribution satisfying
the given constraints, but with a bias toward the distribution $q(x)$,
which, for instance, may contains the information about the range of
the variable $x$. This information, which in the case of the MaxEnt is
not explicitly taken into account, now it is naturally considered from
the beginning. Remarkably, this is a different scenario from that
covered in GUM-S1, i.e., when further information on the quantity is
available, namely, the interval of values within which the
quantity is known to lie is finite.
\par
\begin{figure}[t!]
  \begin{center}
    \includegraphics[width=7cm]{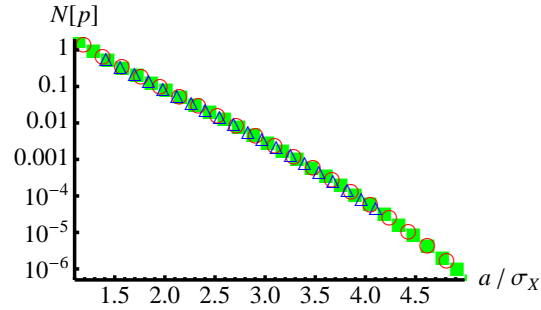}
  \end{center}
  \vspace{-0.3cm}
  \caption{Scaling of the negentropy of mKE distribution for zero mean value 
    and variable known to lie in a symmetric interval $[-a,a]$. 
    We report the negentropy of the distribution as a function of the
    ratio $a/\sigma_\smx$ for different values of the variance:
    $\sigma_\smx^2=0.5$ (green squares), $1$ (red circles), $2$ (blue
    triangles).
    \label{f:f1}}
\end{figure}

Indeed, as mentioned above, if the standard uncertainty is
sufficiently small with respect to the range of variation of the
quantity, we can adopt a Gaussian probability distribution over the
whole real axis and, thus, use the GUM-S1 recommendation. In order to
rigorously justify this statement, which has been qualitatively
addressed in \cite{Lira09}, we assess quantitatively how the knowledge
of the range of variation influences the assignment of a probability
distribution by considering the deviation of the mKE distribution from
a Gaussian distribution, which would represents the MaxEnt solution in
the absence of any information about the range of variation. The
deviation from normality of the mKE distribution (\ref{mKE:sol}) may
be quantified by its negentropy \cite{ng}:
\begin{equation}\label{negE}
  N[p] = \mbox{$\frac12$} \left[1+\log \left(2\pi \sigma^2_\smx\right)\right]
  -S[p]\,,
\end{equation}
where $S[p]$ is the Shannon entropy (\ref{shan}) of the distribution
(\ref{mKE:sol}).  As for example, for a variable known to lie in a
given interval $[a,b] \subset {\mathbbm R}$, $a<b$, that corresponds
to a bias of $p(x)$ toward the flat distribution:
\begin{equation}
  q(x)= \left\{\begin{array}{ll}
      (b-a)^{-1} &\mbox{if $x\in [a,b]$} \\
      0         &\mbox{otherwise}
    \end{array}
  \right. \,,
\end{equation}
the negentropy (\ref{negE}) reads:
\begin{equation}
  N[p] = \mbox{$\frac12$} \left[1+\log \left(2\pi \sigma^2_\smx\right)\right]
  -\log \left(b-a\right) -\lambda_0 - \lambda_1 \bar{x} -\lambda_2
  (\sigma_\smx^2 + \bar{x}^2)\,.
\end{equation}
In the simplest case, namely when $\bar{x}=0$ and $x\in [-a,a]$, the
dependence of the coefficients $\lambda_0$ and $\lambda_2$ is such
that we have a scaling law for negentropy, which depends only on the
ratio $a/\sigma_\smx$. This is illustrated in Fig.~\ref{f:f1}, where
we report the negentropy as a function of $a/\sigma_\smx$ for
different values of $\sigma_\smx$.
%%%%%%%%%%%%%%
\par
In conclusion, we have shown that the determination of the probability
distribution of a variable for which we know the first two moments and
its range of variation may be effectively pursued by using the mKE.
Furthermore, the negentropy of the distribution may be used to
quantify how much the mKE solution differs from the MaxEnt one, i.e.
to assess how the knowledge of the range of variation influences the
assignment of a probability distribution. Our analysis quantitatively
supports the conclusions of Ref.~\cite{Lira09} and rigorously
justifies the use of GUM-S1 recommendation also in the presence of
further information on the range of variation of a quantity, namely,
provided that its standard uncertainty is sufficiently small compared
to the range.
%%%%%%%%%%%%%%
\section*{Acknowledgments}\label{s:ackn}
The authors thank M.~Genovese and I.~P.~Degiovanni for useful discussions.  This work has been supported by MIUR (FIRB ``LiCHIS'' - RBFR10YQ3H), MAE (INQUEST), and the University of Trieste (FRA 2009).
%%%%%%%%%%%%%%

%%%%%%%%%%%%%%

\section*{References}
\begin{thebibliography}{50}
\bibitem{sGUM1} BIPM, IEC, IFCC, ILAC, ISO, IUPAC, IUPAP and OIML
2008 {\it Evaluation of Measurement Data---Supplement 1 to
the Guide to the Expression of Uncertainty in
Measurement---Propagation of distributions using a Monte
Carlo method} Joint Committee for Guides in Metrology,
JCGM 101 \url{http://www.bipm.org/utils/common/documents/jcgm/JCGM_101_2008_E.pdf}
% {\verb
% http://www.bipm.org/utils/common/documents/jcgm/JCGM_101_2008_E.pdf
% }
\bibitem{maxent:57} Jaynes E T 1957 Phys. Rev. {\bf 106} 620;
Jaynes E T 1957 Phys. Rev. {\bf 108} 171
\bibitem{W87} W\"oger W 1987 IEEE Trans. Instr. Measurement {\bf IM-36}
655658
\bibitem{Lira09} Lira I 2009 Metrologia {\bf 46} L27
\bibitem{k1} Kullback S {\em Information theory and statistics}
(Wiley, New York, 1959)
\bibitem{k2} Jaynes E T 1968 IEEE Trans. Systems Science and Cybernetics
{\bf SSC-4} 227
\bibitem{mke1} Olivares S and Paris M G A 2007 Phys. Rev. A {\bf 76}
  042120
\bibitem{ng} Hyvarinen A 1998 Adv. Neural Inf. Proc. Syst. {\bf 10}
  273; Hyvarinen A and Oja E 2000 Neural Networks {\bf 13} 411
\end{thebibliography}
\end{document}